\documentclass[lettersize,journal]{IEEEtran}
\usepackage{amsmath,amsfonts}
\usepackage{algorithmic}
\usepackage{array}
\usepackage{subfigure}
\usepackage{textcomp}
\usepackage{url}
\usepackage{verbatim}
\usepackage{graphicx}
\usepackage{times}
\usepackage{latexsym}
\usepackage{multicol}
\usepackage{xcolor, colortbl}
\usepackage{url,amssymb,threeparttable,multirow,booktabs, tabularx}
\usepackage[sort&compress,square,comma,numbers]{natbib}

\newcommand{\pp}{\textsl{{DynaPose}}}

\newcommand{\cc}{\text{0.984}}
\newcommand{\dt}{\text{0.961}}
\newcolumntype{P}[1]{>{\centering\arraybackslash}m{#1}}

\def\BibTeX{{\rm B\kern-.05em{\sc i\kern-.025em b}\kern-.08em
    T\kern-.1667em\lower.7ex\hbox{E}\kern-.125emX}}
\usepackage{balance}

\begin{document}
\title{Dynamic Anchor Selection and Real-Time Pose Prediction for Ultra-wideband Tagless Gate}

\author{ 
Junyoung Choi, \IEEEmembership{Member,~IEEE},~
Sagnik Bhattacharya, \IEEEmembership{Member,~IEEE},~
Joohyun Lee, \IEEEmembership{Member,~IEEE}
\thanks{
Junyoung Choi is the corresponding author of this paper.
\\
J. Choi, S. Bhattacharya, and J. Lee are with the Communication Standards Research Team at Samsung Research, Samsung Electronics, Seoul, Republic of Korea. Emails: \{juny.choi, sagnik.b, jh5648.lee\}@samsung.com \\
}
} 


\markboth{Journal of \LaTeX\ Class Files,~Vol.~18, No.~9, September~2020}%
{How to Use the IEEEtran \LaTeX \ Templates}

\maketitle

\begin{abstract}
Ultra-wideband (UWB) is emerging as a promising solution that can realize proximity services, such as UWB tagless gate (UTG), thanks to centimeter-level localization accuracy based on two different ranging methods such as downlink time-difference of arrival (DL-TDoA) and double-sided two-way ranging (DS-TWR).
The UTG is a UWB-based proximity service that provides seamless gate pass system without requiring real-time mobile device (MD) tapping.
The location of MD is calculated using DL-TDoA, and the MD communicates with the nearest UTG using DS-TWR to open the gate.
Therefore, the knowledge about exact location of MD is the main challenge of UTG, and hence we provide the solutions for both DL-TDoA and DS-TWR.
In this paper, we propose dynamic anchor selection for extremely accurate DL-TDoA localization and pose prediction for DS-TWR, called {\pp}.
The pose is defined as the actual location of MD on the human body, which affects the localization accuracy.
{\pp} is based on line-of-sight~(LOS) and non-LOS~(NLOS) classification using deep learning for anchor selection and pose prediction.
Deep learning models use UWB channel impulse response and the inertial measurement unit embedded in smartphone. 
{\pp} is implemented on Samsung Galaxy Note20 Ultra and Qorvo UWB board to show the feasibility and applicability.
{\pp} achieves a LOS/NLOS classification accuracy of {\cc}, 62\% higher DL-TDoA localization accuracy, and ultimately detects four different poses with an accuracy of {\dt} in real-time.
\end{abstract}


\section{Introduction}
\label{sec:intro}

Recently, Ultra-wideband (UWB) has emerged as a promising solution for proximity services with centimeter-level localization accuracy compared to other narrowband radios such as Wi-Fi and Bluetooth~\cite{choi2020smartphone, choi2022wand}.
UWB-based proximity services, such as UWB tagless gate (UTG)~\cite{tagless}, UWB payment~\cite{projectnear}, and UWB car key~\cite{ccc, digitalcar} have been developed actively during the last 5 years due to their commercial and social values.
In addition, the flagship smartphones of Samsung (e.g., Galaxy Note 20 Ultra at 2020 and later) are equipped with UWB module.

\begin{figure}[t]
\centering
\includegraphics[width=1\columnwidth]{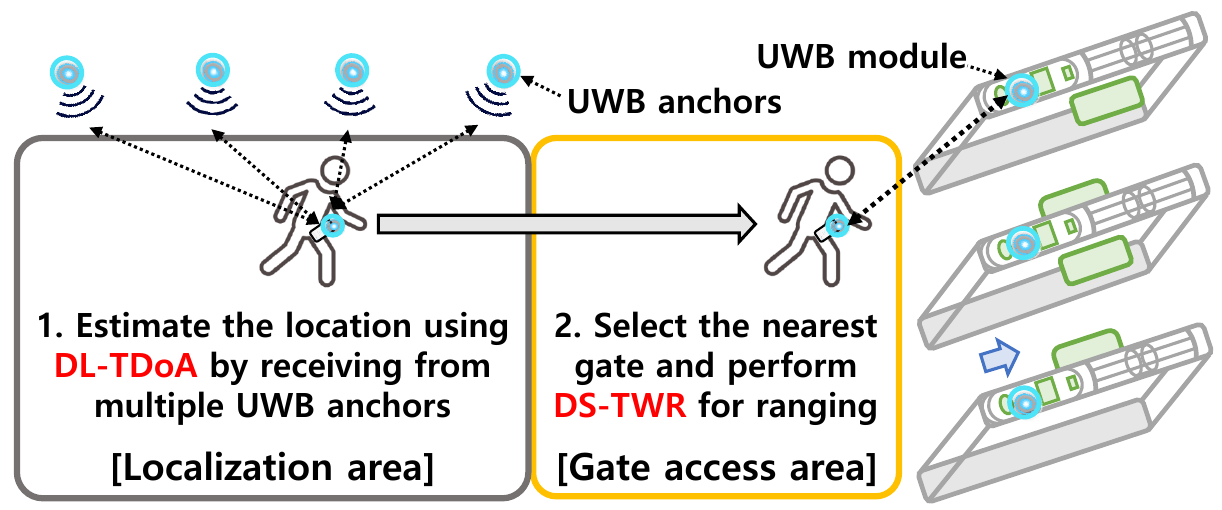}
\caption{The operation of UTG based on the two different UWB ranging technologies.}
\label{fig:utg_operation}
\end{figure}

The aforementioned applications are based on the location of a mobile device (MD) or the distance between two UWB modules, which uses two different ranging methods such as downlink time-difference of arrival (DL-TDoA) and double-sided two-way ranging (DS-TWR). 
For example, UTG uses both DL-TDoA and DS-TWR to provide seamless gate pass without the need to tap the MD on the gate for it to open in the office building or subway station.
The detailed procedure of UTG is illustrated in Fig.~\ref{fig:utg_operation}.
Firstly, the MD calculates its location using timestamps of the received messages from multiple UWB anchors using DL-TDoA in localization area (i.e., gray zone).
When the user enters into the pre-assigned gate access area (i.e., yellow zone), the MD selects the nearest gate and performs DS-TWR with it. Based on the user information exchanged during this DS-TWR process, the gate automatically opens for the user to pass through.
Therefore, the service quality of UTG totally depends on the localization accuracy of DL-TDoA and DS-TWR.

\begin{figure}[t]
    \centering
    \subfigure[Normal]{
    \includegraphics[width=0.31\columnwidth]{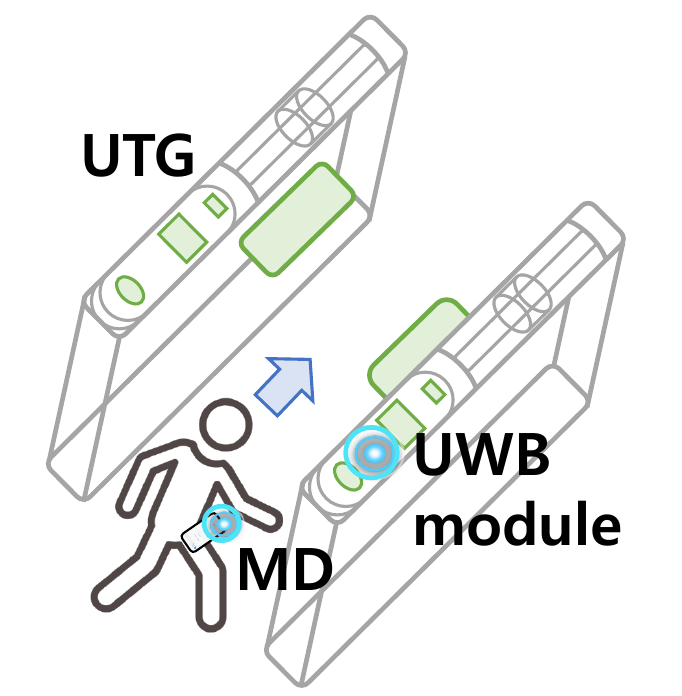}
    \label{fig:normal_operation}
    }
    \subfigure[Unexpected operations]{
    \includegraphics[width=0.61\columnwidth]{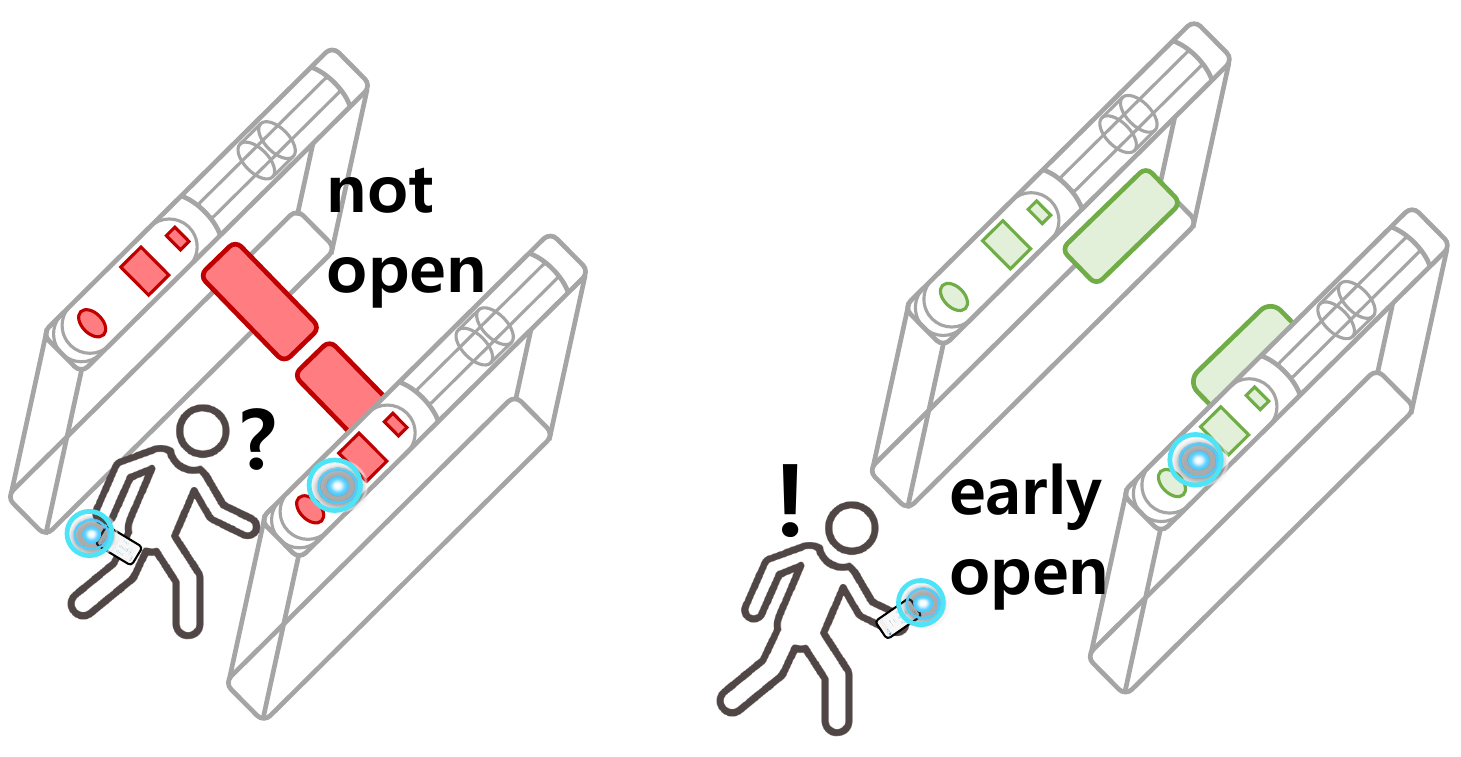}
    \label{fig:unexpected_operation}
    }
    \caption{The normal and unexpected operations of UTG. The opening and closing of gate are colored in green and red.}
    \label{fig:scenario}
\end{figure}

However, even if UWB offers the robust localization performance, it is severely affected by the signal conditions such as line-of-sight (LOS) or non-LOS (NLOS)~\cite{van2021anchor}.
MD needs centimeter-level of localization accuracy during DL-TDoA to perfectly determine whether it is located inside the localization area or gate access area. This necessitates selecting anchors which will provide high anchor-MD channel signal quality.
Inside the gate access area, because of the location of the MD on the user's body, the actual location of the user may be different from that of MD when the user approaches UTG~\cite{grosswindhager2018salma}. In this paper, we define \textit{the position of the MD on the human body} as \textit{pose}. The pose of the MD can lead to the user experiencing unexpected operations, as shown in Fig.~\ref{fig:scenario}.


When the user is close enough to the gate, it must be automatically opened, as shown in Fig~\ref{fig:normal_operation}.
However, if MD is in the back pocket, the gate might not be opened because of the incorrect ranging results due to a blockage, signal attenuation, and multipath, or, the gate might be opened too early when MD is in front of the body, as shown in Fig~\ref{fig:unexpected_operation}.
Due to the difference between the actual location of the human body and the pose of MD, these unexpected operations occur.
These unexpected operations may be resolved when the opening of the gate is adaptively operated based on the pose of MD.

In this paper, we present {\pp}, deep learning-based \textbf{\textsl{Dyna}}mic anchor selection and real-time \textbf{\textsl{Pose}} prediction for UTG. 
We newly design and implement a novel UWB anchor selection and pose prediction system based on LOS/NLOS classification using UWB channel impulse response (CIR) and inertial measurement unit (IMU) sensor of MD.
We define four different poses of MD, and classify them in two steps using deep learning methods, namely convolutional neural network~(CNN) and long-short-term memory (LSTM).
First, {\pp} collects CIR information of UWB messages and determines whether the signal is LOS or NLOS.
When performing LOS/NLOS classification, {\pp} sorts out effective CIR (eCIR) to minimize the input size of the CNN model, and hence, achieves real-time operation without any delay.
Finally, {\pp} determines the subset of anchors providing high signal quality for DL-TDoA localization, and predicts the pose of MD during DS-TWR based on the LOS/NLOS classification.

Our major contributions are summarized as follows:
\begin{itemize}
\item 
To the best of our knowledge, this is the first work to propose real-time LOS/NLOS classification, anchor selection, and pose prediction operating on Android OS with the help of the UWB board.
We believe that our research can be an important technology to encourage UTG proximity services.

\item 
Based on deep learning models, {\pp} is designed by combining the LOS/NLOS classification and pose prediction.
To realize a delay-free real-time operation, only eCIRs are used to classify LOS/NLOS condition to minimize the data transfer latency between the smartphone and UWB board.

\item 
We implemented {\pp} in the commercial Samsung smartphone (i.e., SM-N986N) and Qorvo UWB module (i.e., MDEK1001), and evaluated its performance to show the feasibility.

\item
{\pp} classified LOS/NLOS condition with an accuracy of {\cc}, achieved 62\% higher localization accuracy in NLOS condition, and correctly predicted the four different poses with an accuracy of {\dt} in real-time.
\end{itemize}

\section{Primer: Background}
\label{sec:primer}

\begin{figure}[t]
    \centering
    \subfigure[The procedure of DS-TWR with ranging messages]{
    \includegraphics[width=\columnwidth]{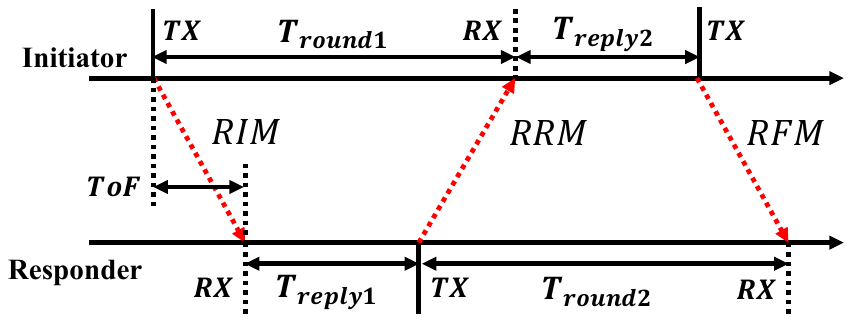}
    \label{fig:dstwr_procedure}
    }
    \subfigure[The procedure of DL-TDoA with multiple anchors]{
    \includegraphics[width=\columnwidth]{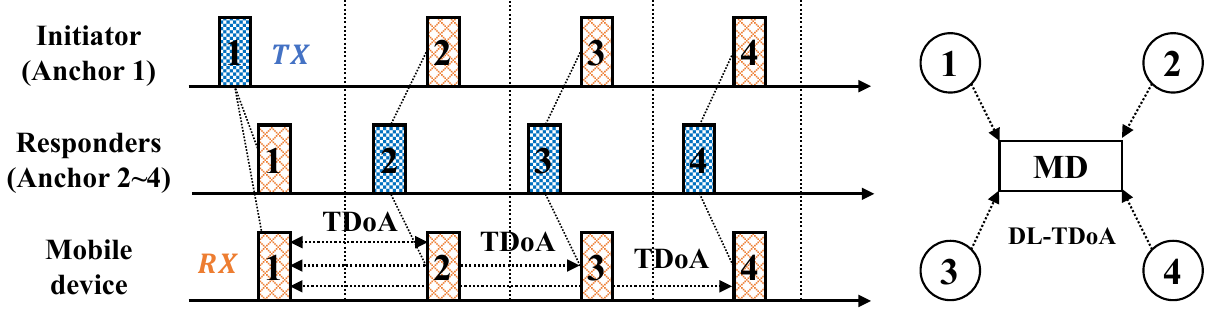}
    \label{fig:dltdoa}
    }
    \caption{Two different ranging methods using UWB.}
    \label{fig:ranging_methods}
\end{figure}

UWB-based wireless communication uses a bandwidth over 500MHz, thus having an ultra-low symbol duration of 2~ns. This high time resolution makes it suitable for ranging and localization operations.

\noindent\textbf{DS-TWR:} The IEEE 802.15.4z standard, published in 2020, defines DS-TWR for ranging to measure the distance between two UWB devices~\cite{ieee80215}. 
DS-TWR is an extension of single-sided two-way ranging in which two round-trip time measurements are used and combined to give the time-of-flight~(ToF) result.
The operation of DS-TWR is shown in Fig.~\ref{fig:dstwr_procedure}.
The initiator measures the first round-trip time by sending ranging initiation message~(RIM) and the responder replies by sending ranging reply message~(RRM).
After sending RRM, the responder measures the second-trip time to which the initiator replies by sending ranging final message~(RFM).
Each device precisely measures the transmission and reception times of the messages, and the resultant of ToF can be estimated by the following equation:
\begin{equation}
    {ToF} = \frac{T_{round1}T_{round2} - T_{reply1}T_{reply2}} {T_{round1} + T_{round2} + T_{reply1} + T_{reply2}}.
\end{equation}

\noindent\textbf{DL-TDoA:} The main concept of this ranging method is that the MD only listens to the messages exchanged by anchors, as shown in Fig.~\ref{fig:dltdoa}.
Anchors exchange messages for time synchronization, and MD measures the time difference of arrival of responder messages based on the message timestamp from the initiator for self-localization.
At least three TDoA measurements with anchor locations are required for localization.
In other words, at least four anchors are needed for operating the DL-TDoA localization system.
Since MD can self-localize by only receiving signals from anchors, no anchor-MD communication is needed.
This makes the DL-TDOA process high scalable with increasing number of MDs in large-scale industrial scenarios. 
For MD to get accurate TDoA measurements, time synchronization between anchors is also important.
Time synchronization is supervised by the initiator anchor by being a time reference.
The rest of the anchors synchronize to the initiator by periodically exchanging messages containing timestamps to measure the distance to the initiator anchor and to use it for time correction.
\section{Data Collection with Experiment}
\label{sec:preliminary}
\subsection{Channel impulse response and signal quality}
\label{subsec:cir_diagnos}
We use Qorvo MDEK1001 equipped with DWM1001 UWB module for our whole experiments to measure CIR data and the channel diagnostics~\cite{mdek1001}.
Both data are collected from the responder whenever MD receives UWB signal during DL-TDoA and DS-TWR. 
To obtain CIR data, we need to access to the register file of Qorvo UWB module and have to calculate the complex value (i.e., a 16-bit real integer and a 16-bit imaginary integer), and this operation brings the latency to get CIR data.
The total 1016 CIRs per signal can be generated from UWB module after calculation\footnote{
The span of CIR data is one symbol time. 
This is 1016 samples for nominal 64~MHz mean pulse repetition frequency.}.
The information about the channel diagnostics includes the first path index (FP\_INDEX), the first path amplitude (FP\_AMPL) and the maximum noise (maxNoise).
The APIs, such as dwt\_readaccdata() and dwt\_readdiagnostics(), are used for CIRs and diagnostics collection. 

\begin{figure}[t]
\centering
\includegraphics[width=\columnwidth]{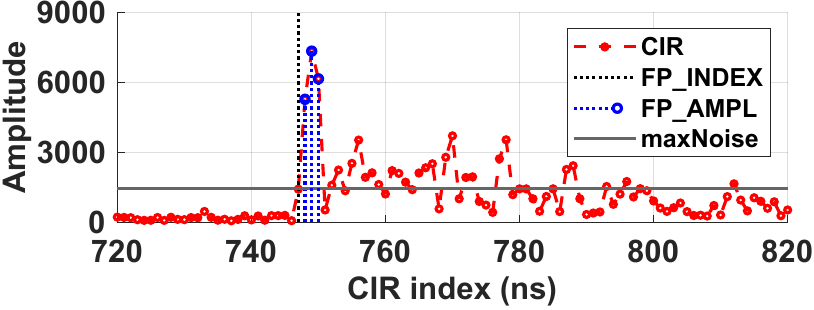}
\caption{The example of CIR data including channel diagnostics. The FP\_INDEX and maxNoise are 747 and 1409.}
\label{fig:cir_example}
\end{figure}

Fig.~\ref{fig:cir_example} shows CIR samples from 720 to 820 and three diagnostics.
The time interval between two consecutive samples is 1~ns.
The black stem, blue stems and horizontal line represent FP\_INDEX, FP\_AMPL and maxNoise, respectively.
The FP\_INDEX gives the actual signal reception timing to help find the start of a meaningful CIR. 
The three consecutive indices after FP\_INDEX are three FP\_AMPLs, which contain the signature of the received signal.
The maxNoise indicates the max amplitude of noise.
We classify the LOS condition by using the above information, and this classification is used for both DL-TDoA and DS-TWR signals.

\begin{figure}[t]
\centering
\includegraphics[width=\columnwidth]{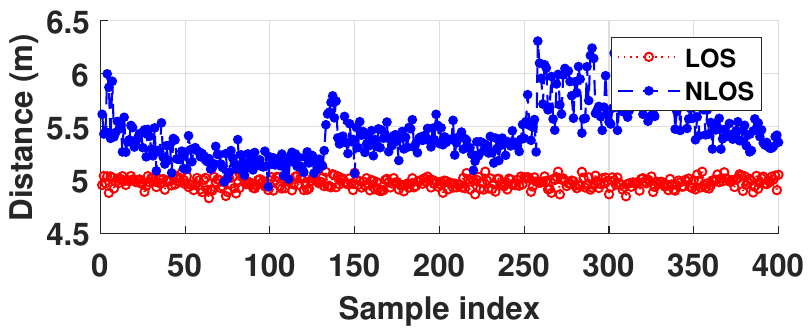}
\caption{The converted distances from TDoA measurements of the LOS (red) and NLOS condition (blue).}
\label{fig:tdoa}
\end{figure}

\subsection{TDoA difference in LOS and NLOS conditions}
\label{subsec:tdoa_diff}
We conducted a preliminary experiment to analyze the effect of LOS and NLOS condition on TDoA measurement.
We installed six UWB anchors (i.e., one initiator and five responders) for DL-TDoA and placed MD in the center of the coverage area of six UWB anchors to generate LOS condition, and blocked the LOS using human body effect to simulate NLOS condition.

Fig.~\ref{fig:tdoa} shows the experiment result of converted distance from TDoA in LOS and NLOS conditions of one anchor among five responder anchors.
LOS condition corresponds to a mean of 497~cm and a standard deviation of 4~cm.
On the other hand, NLOS condition corresponds to a mean of 544~cm and a standard deviation of 26~cm.
The mean difference between LOS and NLOS conditions is 47~cm and the max difference is 133~cm, which is the significant error considering that theoretically UWB can provide centimeter-level localization accuracy.
In addition, the standard deviation also increases up to 22~cm, which might reduce the robustness of UWB localization system.
As a result, we verified that the TDoA measurement in NLOS condition shows the huge mean error and increased standard deviation that harms the reliable localization performance.

\begin{figure}[t]
    \centering
    \subfigure[LOS]{
    \includegraphics[width=0.21\columnwidth]{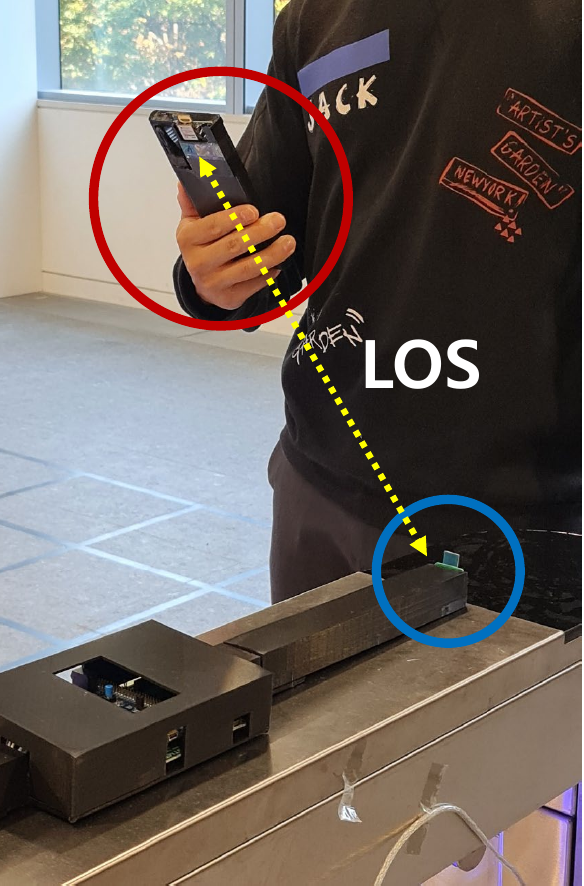}
    \label{fig:handLOS}
    }
    \subfigure[NLOS]{
    \includegraphics[width=0.21\columnwidth]{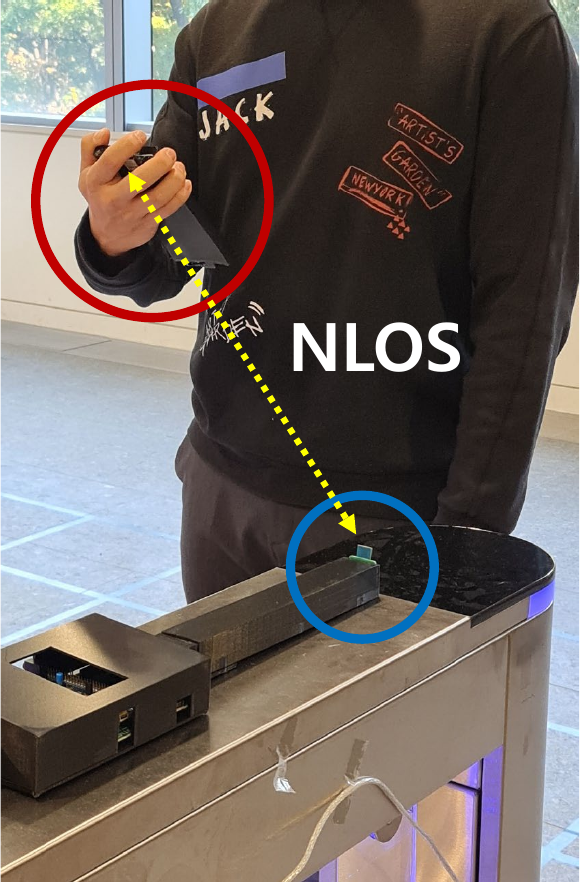}
    \label{fig:handNLOS}
    }
    \subfigure[FRONT]{
    \includegraphics[width=0.21\columnwidth]{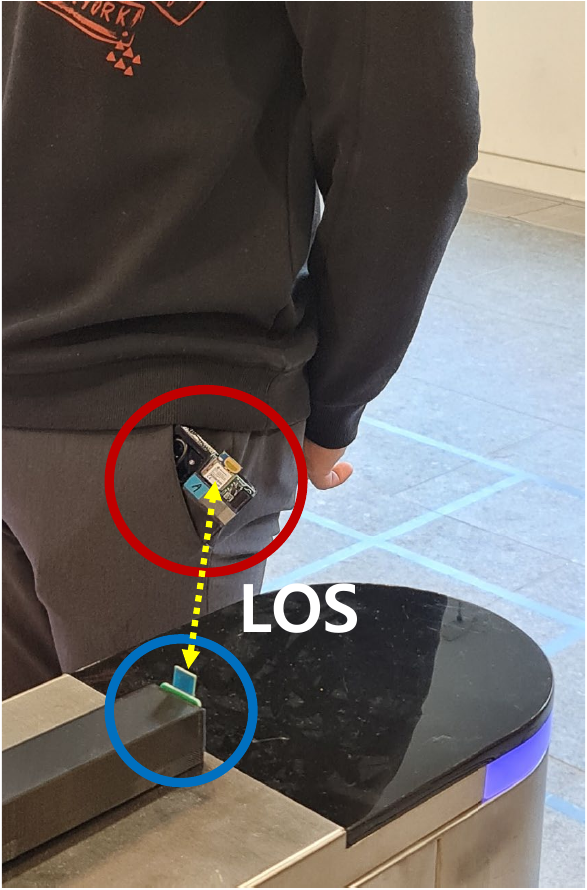}
    \label{fig:frontpocket}
    }
    \subfigure[BACK]{
    \includegraphics[width=0.21\columnwidth]{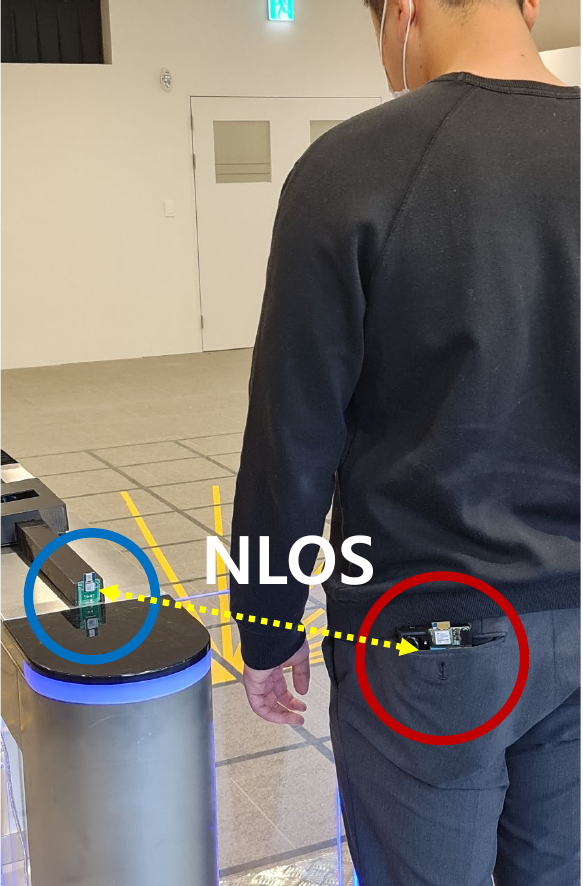}
    \label{fig:backpocket}
    }
    \caption{Snapshots of four different poses for LOS/NLOS.}
    \label{fig:fourpose}
\end{figure}
\begin{figure}[t]
\centering
\includegraphics[width=1\columnwidth]{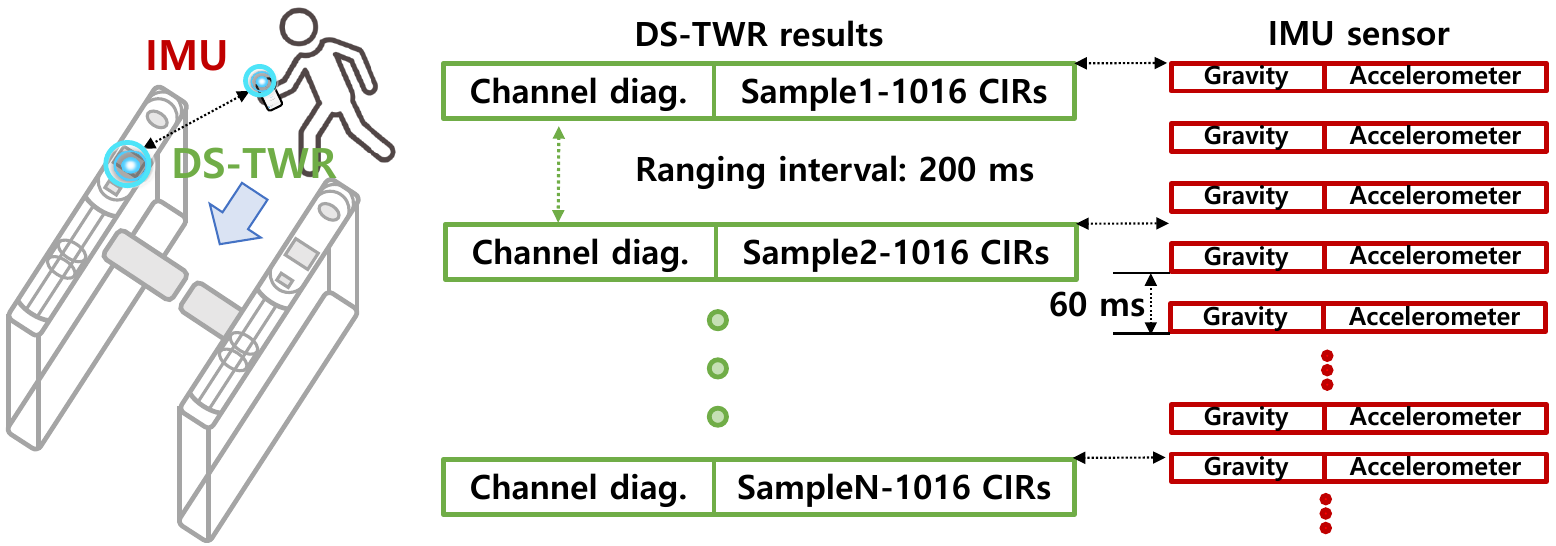}
\caption{The diagnostics and CIR data collection during DS-TWR with IMU. The sampling rates of the ranging results and IMU are 200~ms and 60~ms.}
\label{fig:data_collection}
\end{figure}

\subsection{Definition of four poses of the mobile device}
We adopt a deep learning model for LOS/NLOS classification, and hence, CIR samples with ground truth values of LOS/NLOS are essential. 
In order to collect the LOS/NLOS signals, four representative poses of MD are defined as shown in Fig.~\ref{fig:fourpose}. 
Note that the pose refers to the position of MD on the human body.
We divide the pose into two conditions, such as LOS or NLOS, and both conditions can exist when MD is in the hand or pocket.
Thus, each combination of $\{$LOS,~NLOS$\}\times\{$HAND,~POCKET$\}$ will be one of the poses of MD.
First, ``LOS" is the pose in which the user holds MD without blocking the UWB antenna (Fig.~\ref{fig:handLOS}). 
However, in the case of ``NLOS", the antenna is blocked by hand of user (Fig.~\ref{fig:handNLOS}).
If MD is in the front or back pocket, it is considered as LOS or NLOS environment, respectively.
Since we are considering the scenario that the user is approaching to the gate, MD can receive LOS path directly from the gate if it is in the front pocket, which is called ``FRONT" (Fig.~\ref{fig:frontpocket})\footnote{
We intentionally took a picture of UWB antenna outside the front pocket to clearly show the pose of MD. During the actual experiments, MD is completely inside the front pocket.}.
Conversely, when MD is in the back pocket, the signal becomes NLOS because the human body blocks the signal, which is called ``BACK" ~(Fig.~\ref{fig:backpocket}).

\begin{figure*}[t]
\centering
\includegraphics[width=1\textwidth]{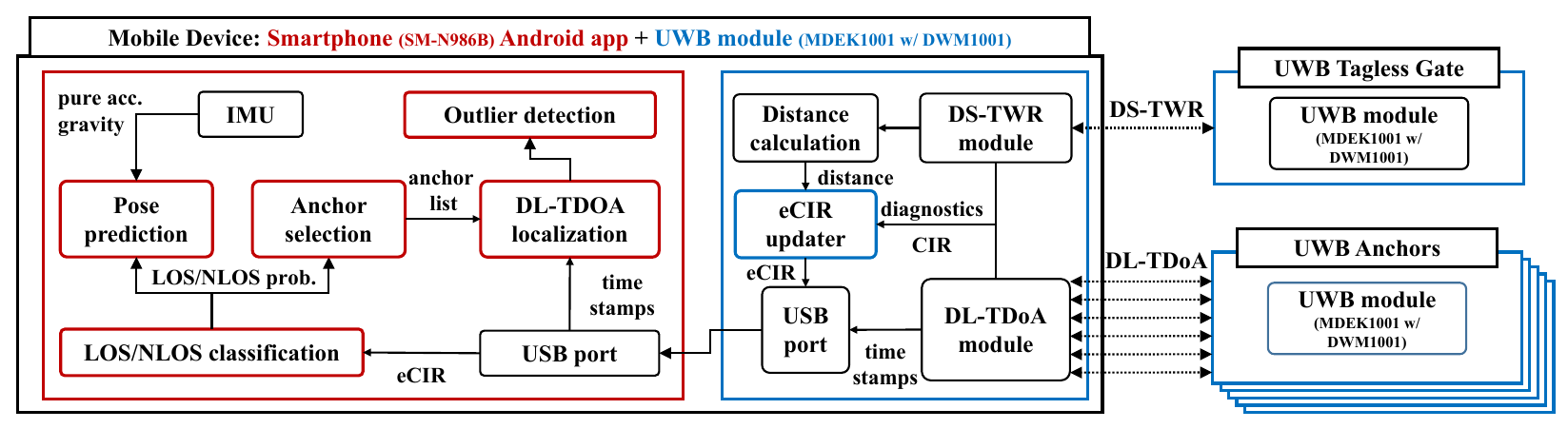}
\caption{The overall architecture of {\pp}. MD is a set of the commercial smartphone and the Qorvo UWB board using wired connection.}
\label{fig:architecture}
\end{figure*}

\subsection{CIR and IMU data collection}
In order to train LOS/NLOS classification model and pose prediction model, CIR data with LOS/NLOS labels and IMU data with pose labels are required.
Based on the four different poses of MD presented in Fig.~\ref{fig:fourpose}, CIR and IMU data samples are collected through preliminary experiments.
Fig.~\ref{fig:data_collection} shows the concept of data collection.
We assume that the user gets close to the gate while MD and the gate perform DS-TWR.
The ranging interval is set to as the default value of 200~ms. 
Therefore, the MD collects the ranging results (i.e.,  1016 CIRs and a channel diagnostics) once every 200~ms.
We collect 2,000 DS-TWR results for each pose, which means that 4,000 LOS and 4,000 NLOS samples are collected for subsequent training.
We also include an open source dataset to prevent the overfitting problem associated with data shortage~\cite{bregar2016nlos}. 
The open dataset contains 42,000 samples and this LOS/NLOS open datasets were created by using Qorvo DWM1001 UWB radio module. Measurements were taken on 7 different indoor locations. In every indoor location 3,000 LOS and 3,000 NLOS samples were collected.
Therefore, a total of 50,000 samples are used for training and testing the CNN-based model.

In addition to this, the IMU sensor data is simultaneously collected once every 60~ms.
The IMU sensor data is composed of the gravity and a pure acceleration of x,y, and z coordinates, which means that 6 sensor values are collected at each timestep.
The reason that we collect those two type of data is that the gravity values contain the information about the orientation of the smartphone and the accelerometer values reflect the signature of the user movement, especially during walking~\cite{shu2015last}.
Since IMU data resolution is approximately 3 times higher than DS-TWR, we collect 6,600 IMU samples for each pose.
In case of IMU, no open source dataset is used, and hence, in total 26,400 samples are used.

\section{{\pp}: Proposed System}
\label{sec:proposed}

\subsection{Overview}
\label{subsec:overview}
Fig.~\ref{fig:architecture} shows the overall architecture of {\pp}.
The MD comprises the smartphone and the Qorvo UWB board (i.e., MDEK1001)  attached through USB serial cable.
The latest smartphones from Samsung, such as Galaxy S23+ and Galaxy Z Fold4, have embedded UWB modules, but there is no developer API to obtain CIR yet. 
This is the reason why we choose 3rd party UWB module for implementation.
The UTG is also equipped with the same UWB module of MD to perform UWB ranging.
UWB module on MD collects CIRs and channel diagnostics of the received messages during DL-TDoA and DS-TWR, and sends these information to eCIR updater.
The eCIR updater generates eCIR from 1016 CIRs based on channel diagnostics such as FP\_INDEX and maxNoise (Section~\ref{subsec:eCIR}).
After eCIR is generated, it is transferred from Qorvo board to the smartphone through USB port.
The smartphone runs the LOS/NLOS classification to determine whether the received signal is LOS or NLOS whenever eCIR reception occurs (Section~\ref{subsec:los_nlos_classifier}).
The LOS/NLOS classification sends the result of the classification to the anchor selection and the pose prediction.
During DL-TDoA in localization area, anchors are selected based on the NLOS probability with Graham Scan algorithm (Section~\ref{subsec:Anchor_selection}), and localization is performed by using only selected anchors.
The outlier detection checks the result of localization to determine whether it is an outlier or not (Section~\ref{subsec:outlier_detection}).
Finally, the pose prediction determines the pose of MD by combining IMU sensor data during DS-TWR in gate access area, especially accelerometer and gravity measurements (Section~\ref{subsec:pose_prediction}).
In both MD and UTG, each process of all the components operates in real-time.

\subsection{Effective channel impulse response updater}
\label{subsec:eCIR}

Since {\pp} performs LOS/NLOS classification, anchor selection and the pose prediction of MD on Android application, the data should be transferred from the Qorvo board to the smartphone through USB port. 
Therefore, the latency of data transfer between the Qorvo board and the smartphone should be optimized to achieve real-time operation.
If the processing time of the whole procedure from Qorvo board to smartphone is larger than the ranging interval, the ranging operation between Qorvo board of MD and UTG will be interrupted.

\begin{figure}[t]
    \centering
    \subfigure[The CIR divided into signal and noise based on maxNoise]{
    \includegraphics[width=0.435\columnwidth]{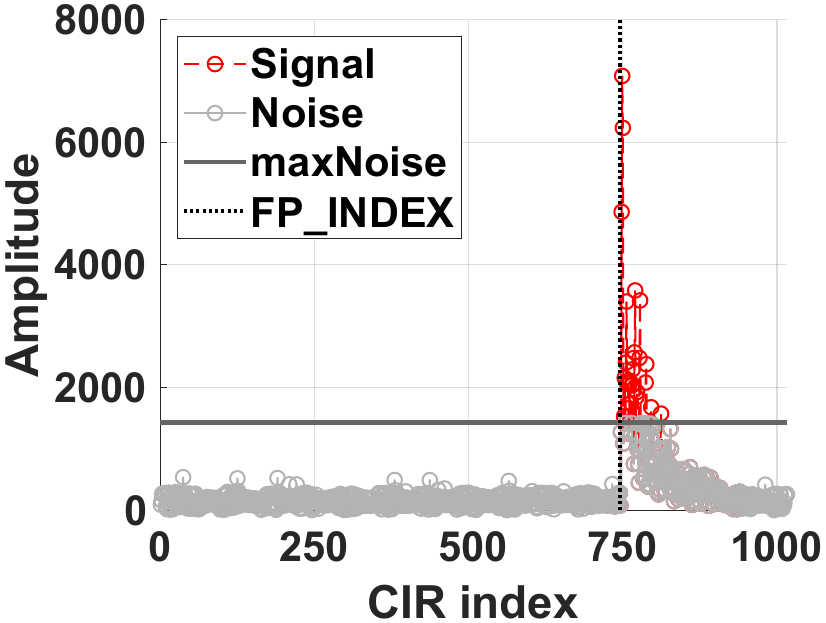}
    \label{fig:cir1016}
    }
    \subfigure[The difference in the amplitude and the number of peaks]{
    \includegraphics[width=0.435\columnwidth]{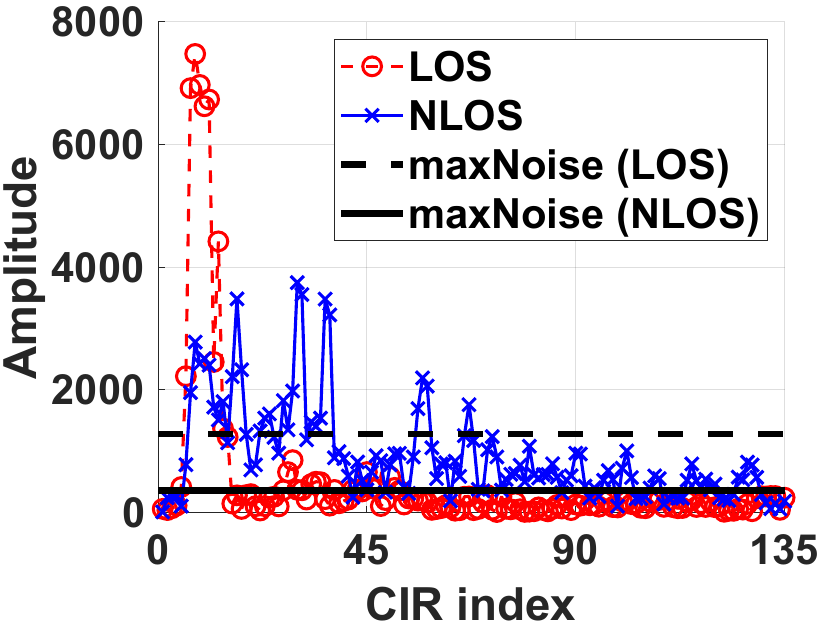}
    \label{fig:cir200}
    }
    \caption{The example of effective CIR and difference between LOS/NLOS.}
    \label{fig:eCIR}
\end{figure}

First, we experimentally measured the latency when three channel diagnostics and 1016 CIRs are transferred to the smartphone through USB without data loss, and 223.4~ms is consumed on average, which is longer than the default ranging interval.
This latency contains the delay of complex value calculation as explained in Section.~\ref{subsec:cir_diagnos}.
Therefore, the latency should be minimized by selective data transfer, and hence, we adopt the eCIR updater to collect only effective CIRs.
Fig.~\ref{fig:cir1016} represents the CIRs divided into a real signal and a noise using maxNoise.
If the amplitude of CIR is less than maxNoise, the CIR is considered as noise.
Thus, CIRs before FP\_INDEX are noise and can be eliminated.
Therefore, we focus on CIRs after FP\_INDEX, which contains the real information of the signal.

Fig.~\ref{fig:cir200} shows only 135 CIRs of two LOS and NLOS CIR samples.
The 0 index represents the index of FP\_INDEX-5. 
Including these 5 extra samples prior to FP\_INDEX serves as a margin for FP\_INDEX in both CIR samples. 
The CIR trends have clear differences between LOS and NLOS in the max amplitude of the peak and the presence of multiple peaks after the first peak.
Those signatures represent the nature of LOS and NLOS signals.
In the case of LOS, the amplitude of CIRs quickly goes below maxNoise, and, only 61 CIRs are actual meaningful signals after FP\_INDEX in average.
However, the first peak amplitude of the NLOS signal is much smaller than that of LOS, and multiple peaks are observed after the first peak because of the multipath signals in NLOS situation.
Therefore, on an average, 130 CIRs are larger than maxNoise after FP\_INDEX in case of NLOS.
As a result, at least 135 CIRs (i.e., 130 and 5 margin CIRs before FP\_INDEX) should be transferred to the smartphone for LOS/NLOS classification since those CIRs contain all the signatures of the received signal.
We consider those 135 CIRs as eCIR and use only 135 CIRs for LOS/NLOS classification instead of 1016 CIRs.
As a result, the latency is reduced to 17.8~ms from 223.4~ms, which is much smaller than the ranging interval.
 
\subsection{LOS/NLOS classification}
\label{subsec:los_nlos_classifier}

\begin{figure}[t]
    \centering
    \subfigure[LOS/NLOS classification model architecture with 4 CNN layers]{
    \includegraphics[width=\columnwidth]{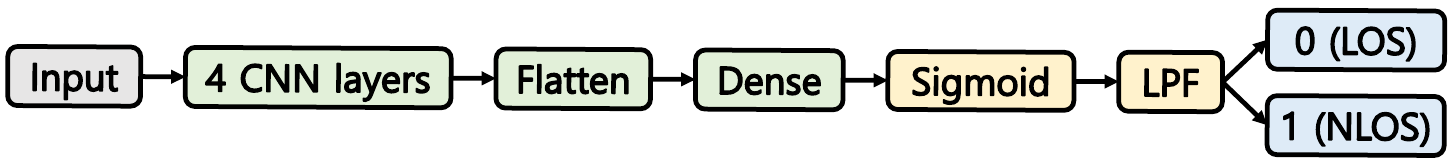}
    \label{fig:nlos_classifier_architecture}
    }
    \subfigure[CNN layer architecture]{
    \includegraphics[width=\columnwidth]{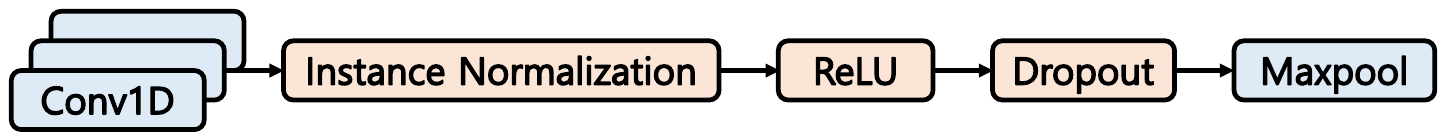}
    \label{fig:nlos_classifier_cnn}
    }
    \caption{CNN based model for LOS/NLOS classification using eCIR input.}
    \label{fig:nlos_classifier}
\end{figure}

Fig.~\ref{fig:nlos_classifier_architecture} shows the deep learning model architecture adopted for the CIR based LOS/NLOS classification, which is applied to both DL-TDoA and DS-TWR messages.
The input layer comprises the 135 consecutive CIRs from FP\_INDEX-5. 
It is followed by 4 CNN layers, which are highly efficient in extracting meaningful practical features from spatial/temporal data. 
The output of the CNN is then flattened from the 2D image format to a 1D vector, using the flatten layer. 
This is followed by a dense layer, in which all the neurons of the previous and the current hidden layer are pairwise connected using weights. 
The output of the dense layer is passed into the sigmoid activation layer, which applies sigmoid function to produce a number between 0 and 1, which in this case indicates the probability of the CIR corresponding to a NLOS condition. 
Finally, the output of the sigmoid is smoothed by the low-pass filter~(LPF) to remove the outliers, and the signal is classified as 0 (LOS) or 1 (NLOS).

\begin{table}[t]
\centering
    \caption{The LOS/NLOS classification model parameters}
    \begin{tabular}{c|c}
    \toprule[0.6pt]\midrule[0.3pt]
    Model parameters & Value \\
    \midrule
    Conv1D layer [1,2,3,4] kernel size & [5,11,17,5]  \\
    Conv1D layer [1,2,3,4] \# of filters & [64,128,256,512] \\
    Dropout rate & 0.2 \\
    Maxpool layer kernel size & 2 \\
    Optimizer & Adam \\
    Loss function & Binary cross-entropy \\
    Batch size & 50 \\
    Maximum no. of epochs & 20 \\
    \midrule[0.3pt]\bottomrule[0.6pt]
  \end{tabular}
  \label{tab:NLOS}
\end{table}

The detailed architecture of an individual CNN layer is shown in Fig.~\ref{fig:nlos_classifier_cnn}. 
The convolution layer Conv1D consists of a pre-specified number of 1D vectors called filters, all of them having a pre-specified size called kernel size. 
These filters help to extract significant spatial/temporal features from the input data. Each of these filters is convolved with the input data to produce a vector. 
These output vectors are stacked one above another to give a 2D image type output. 
The output of the convolution layer is passed into the instance normalization~(IN) layer, which applies normalization per data sample along a specified axis. 
Unlike the conventional batch normalization layer, IN reduces incoming noise and model dependence on the training data statistics. 
The normalized output is sent into the rectified linear units~(ReLU) layer, which applies the ReLU activation function.
This is followed by the dropout layer, which randomly drops a fraction of the neurons at algorithm runtime. 
This reduces the overfitting error of the model to the training data. 
The final layer is a maxpool layer, which also operates based on a pre-specified kernel size. 
It performs the maxpooling operation (i.e. chooses the maximum element along a specified axis) in the pre-specified kernel size, and replaces the elements in that kernel with this chosen maximum element. 
This helps to bring down the dimensionality of the data and helps to simplify the model and prevent overfitting. 
The entire set of parameters used in the CNN based model for the LOS/NLOS classification is listed in Table~\ref{tab:NLOS}.

\begin{figure}[t]
\centering
\includegraphics[width=1\columnwidth]{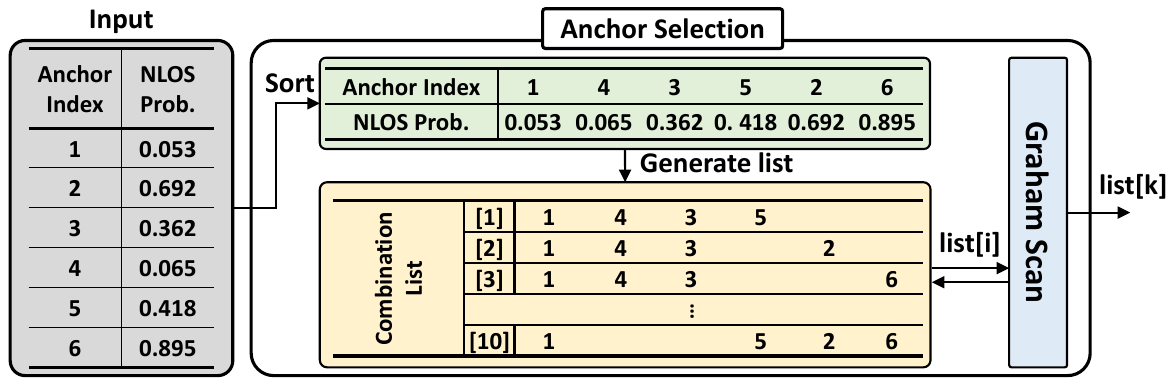}
\caption{The anchor selection example when $n$ is 6.}
\label{fig:anchor_selection}
\end{figure}

\subsection{Anchor selection algorithm}
\label{subsec:Anchor_selection}

{\pp} activates anchor selection algorithm~(ASA) when at least one anchor-MD path is in NLOS condition. 
In other words, {\pp} deactivates ASA when all anchors correspond to LOS conditions.
Once ASA is activated, ASA iterates to find out the best combination of 4 anchors that satisfies the following two conditions: 1) low NLOS probability and 2) convex hull of anchors containing MD inside\footnote{ASA selects 4 anchors when it is activated, which is a minimum number of anchors for TDoA localization.}.
Note that the NLOS condition anchor has a biased mean and high standard deviation of TDoA compared to the anchor with the LOS condition as described in Section~\ref{subsec:tdoa_diff}.
In addition, the localization using TDoA suffers from a convex hull problem, and localization accuracy can be negatively affected if MD is located outside of the convex hull formed by anchors.
Hence, it is important to select an anchor combination that embraces MD in a convex hull.
Consequently, we apply ASA satisfying two conditions to minimize the localization error caused by NLOS condition anchors and convex hull problem.

The detailed operation of ASA is illustrated in Fig.~\ref{fig:anchor_selection}.
ASA first sorts anchors in ascending order of NLOS probability.
ASA arranges all cases of $_\textrm{n}$\textbf{C}$_4$ based on the sorted list with four selected anchors among total $n$ anchors.
The list starts from the combination of the first four anchors, which shows the four lowest NLOS probabilities.
Sorting the anchors can naturally give priority to the anchors with lower NLOS probability.
Since one initiator among $n$ anchors is used as a reference for computing TDoA values as explained in Section~\ref{sec:primer}, the initiator anchor should be always included in the anchor combination.
Therefore, the cases without initiator anchor are excluded from the combination list.

Starting from the first case of combination list, ASA performs Graham Scan algorithm~\cite{graham1972efficient} to check whether anchor combination can envelop MD in convex hull.
Graham Scan is an algorithm that finds a convex hull for a given list of coordinates in 2D plane.
Using the coordinates of anchors and MD as input data, Graham Scan outputs the set of vertices on the convex hull.
The input for the coordinate of MD is the previous location during DL-TDoA because the current location of MD is not determined yet.
If MD is not included in the set (i.e., all vertices of the convex hull consist of anchors), we can determine MD is in the convex hull of anchors, and hence, the corresponding case is selected and passed to DL-TDOA localization.
On the other hand, if MD is included in the set, it indicates that MD is located outside of the convex hull of anchors and the next case of combination list is fed to Graham Scan.
ASA sequentially computes the cases of the anchor combination until it finds a case that meets the condition.

\subsection{Outlier detection}
\label{subsec:outlier_detection}
After ASA, {\pp} calculates the location of MD using DL-TDoA (i.e., instantaneous location) and lastly checks whether the location is an outlier or not.
We conceptually use the distance threshold to eliminate the outlier. 
If the distance between the instant and average location is lower than the threshold, it is determined as valid location and passes the outlier detection.
On the other hand, if the distance is higher than the threshold, that instant location is considered as an outlier and is discarded.
The main purpose of outlier detection is to filter the error generated by the initiator when the signal condition is NLOS.
Since ASA should always include the initiator anchor, the outlier could be generated depending on the NLOS condition of the initiator.


There are two data structures to perform outlier detection: data stream buffer (DSB) and inlier (i.e., the opposite of outlier) data buffer.
The DSB enqueues every input data without filtering and the inlier data buffer (IDB) only enqueues the input data when it is determined as an inlier.
Both of the data structure has the same buffer size and they dequeue the oldest data when new data is enqueued.
The reason is to keep the buffer size fixed and also to follow the current trend.
Two buffers basically enqueue the same data except that IDB only enqueues the inlier data.
The IDB is used to classify whether input data is an outlier or not, and DSB is used to monitor the movement of MD so that outlier detection does not misjudge walking movement as an outlier.
This cross-check method prevents misjudging walking movements as outliers.
As a result, {\pp} can calculate the location with extremely high accuracy.



%
\begin{figure}[t]
    \centering
    \subfigure[LOS condition]{
    \includegraphics[width=0.46\columnwidth]{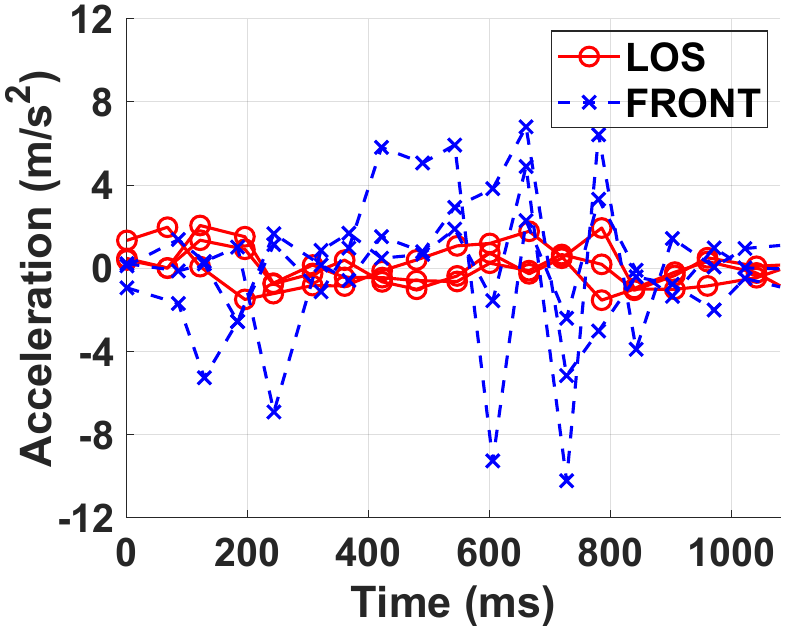}
    \label{fig:imu_los}
    }
    \subfigure[NLOS condition]{
    \includegraphics[width=0.46\columnwidth]{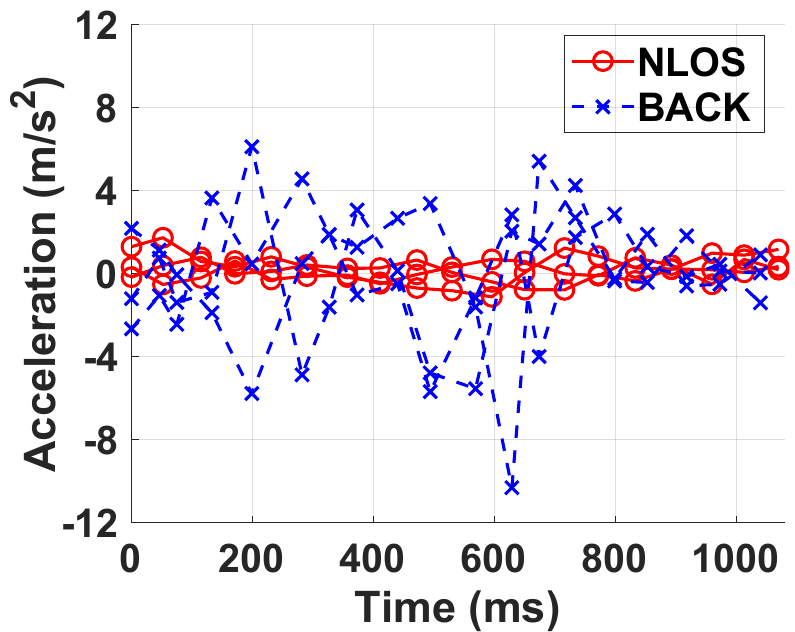}
    \label{fig:imu_nlos}
    }
    \caption{The snapshots of 18 accelerometer values (1080~ms) of four poses.}
    \label{fig:imu_diff}
\end{figure}

\subsection{Pose prediction}
\label{subsec:pose_prediction}

Fig.~\ref{fig:imu_diff} shows the snapshots of IMU sensor values, especially the accelerometer readings of each pose, divided into LOS and NLOS conditions. 
The red and blue plot represent the poses of MD, respectively, and three lines of each pose represent the acceleration value of x,y, and z coordinates.
The trends of acceleration value show clear difference whether pose is HAND or POCKET.
If the pose of MD is close to the ground such as POCKET, the magnitude of acceleration is much higher than that of HAND case.
Therefore, the difference in acceleration is a signature of HAND and POCKET, and can be classified by deep neural network.

\begin{figure}[t]
    \centering
    \subfigure[Pose prediction model architecture with 3 CNN layers]{
    \includegraphics[width=\columnwidth]{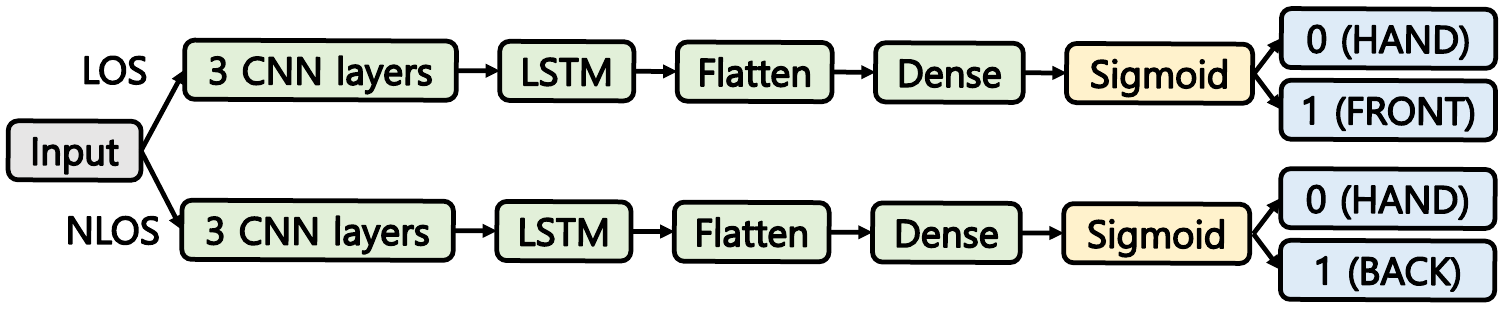}
    \label{fig:pose_classifier_architecture}
    }
    \subfigure[CNN layer architecture]{
    \includegraphics[width=\columnwidth]{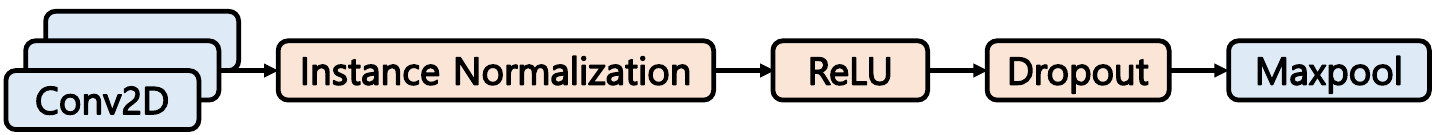}
    \label{fig:pose_classifier_cnn}
    }
    \caption{CNN-LSTM based model for pose prediction using IMU based on LOS/NLOS classification model output.}
    \label{fig:pose_classifier}
\end{figure}

Fig.~\ref{fig:pose_classifier_architecture} shows the deep neural network model used for pose classification. 
The input layer comprises the IMU sensor data with 6 features, namely the accelerometer and gravity sensor readings along x,y, and z coordinates. 
We also accumulate this IMU sensor data for the previous 18 timesteps, which leads to the input being a matrix of dimensions $18\times6$. 
This number of timesteps is derived based on the logic that we want to send the sequential data of the previous 2-3 steps of the user as input data. 
Considering that the average human walking speed is 3 steps per second, that corresponds to sending 1~s worth of IMU sensor data~\cite{shu2015last}. 
The IMU sensor collects data every 60~ms, hence, to collect the IMU data for 1~s, we need to collect 18 IMU data samples. 

The IMU input data is passed into one of two CNN-LSTM models based on whether the CIR based LOS/NLOS classification in the previous stage produced an output LOS or NLOS. 
The two CNN-LSTM based models are designed identically but get trained with different sets of training data, resulting in different optimal weights and filters. 
The model starts with 3 CNN layers, followed by the LSTM layer. 
While the CNN layers extract meaningful spatial/temporal features from the IMU data, the LSTM layer takes as input the time series (18 timesteps) data to produce predictions based on current timestep as well as the previous timesteps~\cite{hochreiter1997long}.
The LSTM output is flattened using a flatten layer, and then input to a dense layer. 
Finally, sigmoid activation layer operates on the previous layer output, to produce a number between 0 and 1. 
The output of sigmoid is directly used for pose prediction without passing through LPF because LOS/NLOS classification results are already filtered by LPF.

For the model corresponding to the LOS case, the output of sigmoid corresponds to the probability of MD being FRONT, as opposed to being LOS with hand.
Similarly, for the NLOS case, the output of sigmoid corresponds to the probability of MD being BACK, as opposed to being NLOS with hand.
The detailed CNN layer architecture used for the pose prediction models is shown in Fig.~\ref{fig:pose_classifier_cnn}. 
The convolution layer Conv2D contains 2D filters of pre-specified kernel size. 
The output of the Conv2D layer is successively passed through an IN layer, ReLU activation, dropout, and a maxpool layer. 
The entire set of parameters used in CNN-LSTM based model for the pose prediction are listed in Table~\ref{tab:pose}.

\begin{table}[t]
\centering
    \caption{The set of pose prediction model parameters}
    \begin{tabular}{c|c}
    \toprule[0.6pt]\midrule[0.3pt]
    Model parameters & Value \\
    \midrule
    Conv2D layer [1,2,3] kernel size & [2,2,2]  \\
    Conv2D layer [1,2,3] \# of filters & [64,128,256] \\
    Dropout rate & 0.2 \\
    Maxpool layer kernel size & 2 \\
    LSTM layer no. of units & 128 \\
    Optimizer & Adam \\
    Loss function & Binary cross-entropy \\
    Batch size & 100 \\
    Maximum no. of epochs & 100 \\
    \midrule[0.3pt]\bottomrule[0.6pt]
  \end{tabular}
  \label{tab:pose}
\end{table}

\section{Performance Evaluation}
\label{sec:evaluation}

\subsection{Implementation}
\label{subsec:implementation}
\noindent\textbf{MD:}
We implemented the application side of {\pp} on the commercial smartphone (i.e., SM-N986B) running on Android~11 to show the feasibility.
All modules, such as LOS/NLOS classification, anchor selection, outlier detection, and pose prediction, are implemented using Android Studio.
Both LOS/NLOS classification and pose prediction that includes CNN models are implemented using TensorFlow 2.10.0 and converted to TensorFlow-Lite for mobile deployment.
For LPF included in LOS/NLOS classification, the exponentially weighted moving average is used with a weighting factor of 0.8.
If the user turns on {\pp}, the results of LOS/NLOS classification and pose prediction are displayed on application UI.

\noindent\textbf{UWB:}
We implemented three operating components of {\pp}, such as UWB module on MD, UWB anchor, and UTG, based on the example code from the Qorvo DWM1001~\cite{dwm1001example}. 
The DL-TDoA operation between MD and anchor, DS-TWR operation between MD/UTG, and eCIR updater in Qorvo board is implemented by following the user manual using SEGGER Embedded Studio for ARM 5.32a~\cite{dw1000user, seggerARM}.
The UWB module is connected to MD via a wired connection.

\subsection{Experiment setup}
\label{subsec:experiment}
\noindent\textbf{Datasets and CNN training:} We train our LOS/NLOS classification on open datasets and collected data with preliminary experiments to prevent overfitting, and pose prediction is trained by using only collected IMU data.
All datasets were randomly divided into training (80\%) and testing (20\%) data.

\begin{figure}[t]
\centering
    \includegraphics[width=\columnwidth]{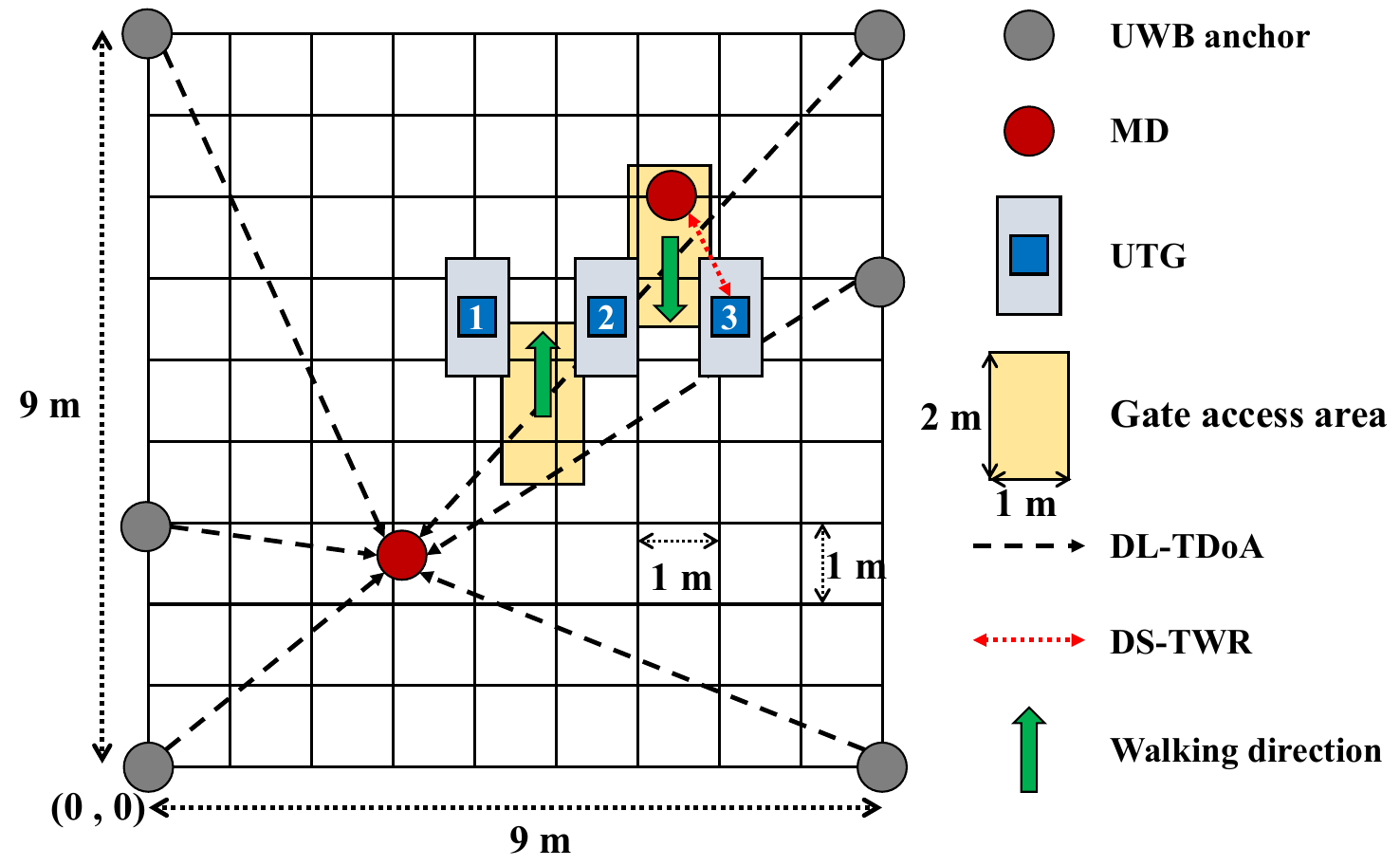}
\caption{The illustration of experiment environment.}
\label{fig:deploy_blueprint}
\end{figure}

\begin{figure}[t]
\centering
\includegraphics[width=\columnwidth]{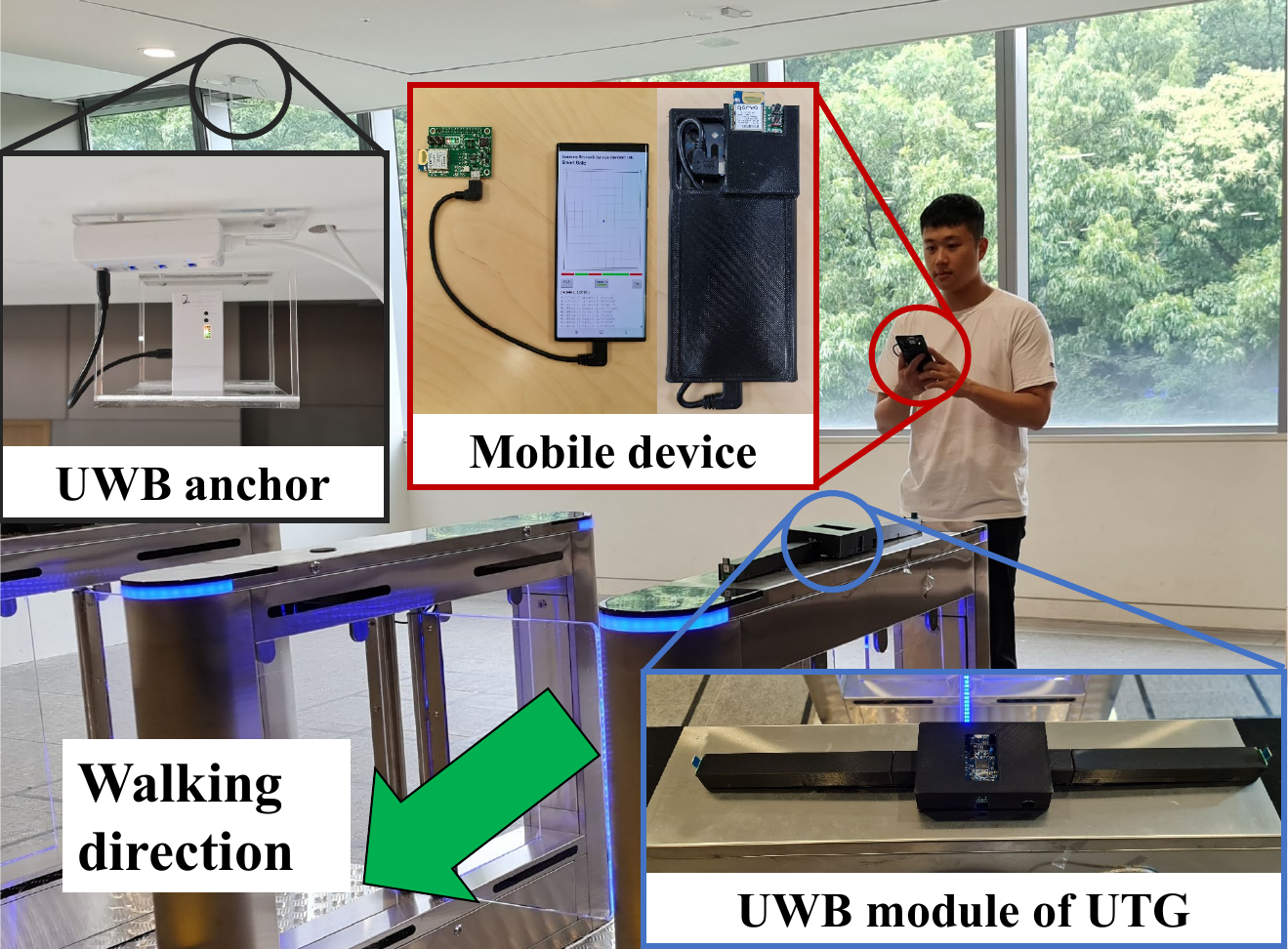}
\caption{The snapshot of the real-world UTG experiment.}
\label{fig:real-world}
\end{figure}

\noindent\textbf{Experiment environment:}
We conducted our experiments for a test area of $9\times9~m^2$ in an office building with six UWB anchors and UTG as shown in Fig.~\ref{fig:deploy_blueprint}.
We set the gate access area as $1\times2~m^2$.
UTG is installed with UWB module and MD is composed of the smartphone and UWB module, and we also designed the case of MD using a 3D printer to fasten UWB module for the consistency of experiments as shown in Fig.~\ref{fig:real-world}.
The user stands inside the gate access area between UTG 2 and 3. 
The experiments are focused to show the real-time accuracy of LOS/NLOS classification, DL-TDoA localization performance, and pose prediction.
The ranging frequency of DL-TDoA and DS-TWR are 500~ms and 200~ms, respectively.
Each experiment contains the result of localization and pose prediction while the user with MD starts walking toward UTG from 3~m away, and we iterate the experiment 100 times.

For easy understanding, we first explain the performance of pose prediction with LOS/NLOS classification accuracy, and localization performance of DL-TDoA with anchor selection will be followed.

\subsection{Performance of pose prediction}
\label{subsec:performance_pose}
We measure the LOS/NLOS classification and pose prediction performance as the rate of true positive and false positive, in which the classification and prediction are correctly/wrongly conducted. 
Table~\ref{tab:performance} summarizes the accuracy of each pose measured in real-time.

\noindent\textbf{LOS/NLOS classification:}
The two values are obtained from experiments.
The first value is the actual performance of LOS/NLOS classification model and the second value in the parentheses represents the accuracy without LPF, which is the output of sigmoid in Fig.~\ref{fig:nlos_classifier_architecture}.
The performance without LPF is similar to that of \cite{jiang2020los}, which proposes CNN-LSTM model for classification, and overall accuracy of 0.819.
{\pp} improves the accuracy by adding LPF to CNN model, and the overall LOS/NLOS classification accuracy is 0.984.
As a result, {\pp} improves classification performance by 20.1\%.

\begin{table}[t]
\centering
    \caption{The real-time performance of LOS/NLOS classification and pose prediction accuracy.}
    \begin{tabular}{c|c|c}
    \toprule[0.6pt]\midrule[0.3pt]
     & {LOS/NLOS classification (w/o LPF)} & {Pose prediction} \\
    \midrule
    {LOS} & 0.994~~(0.849) & 0.983 \\
        \midrule
    {NLOS} & 0.993~~(0.839) & 0.982 \\
        \midrule
    {FRONT} & 0.968~~(0.783) & 0.931 \\
        \midrule
    {BACK} & 0.982~~(0.803) & 0.948 \\
    \midrule[0.3pt]\bottomrule[0.6pt]
  \end{tabular}
\label{tab:performance}
\end{table}

\begin{table}[t]
\centering
    \caption{The transition delay between LOS and NLOS conditions in ms.}
    \begin{tabular}{c|c|c}
    \toprule[0.6pt]\midrule[0.3pt]
     & LOS & FRONT\\
    \midrule
    {NLOS} & 734.8~~(137.4) & 838.4~~(168.9) \\
    \midrule
    {BACK} & 848.0~~(173.8) & 851.2~~(188.4) \\
    \midrule[0.3pt]\bottomrule[0.6pt]
  \end{tabular}
\label{tab:delay}
\end{table}

\noindent\textbf{Pose prediction:}
Considering that the accuracy of pose prediction is affected by the result of LOS/NLOS classification, the pose prediction accuracy can not outperform LOS/NLOS classification accuracy.
The average performance of pose prediction is 0.961, and there is only accuracy degradation of 0.023 compared to LOS/NLOS classification.
Therefore, we conclude that {\pp} successfully classifies four different poses based on the eCIR and IMU.

\begin{figure*}[!t]
    \centering
    \subfigure[LOS condition]{
    \includegraphics[width=0.24\textwidth]{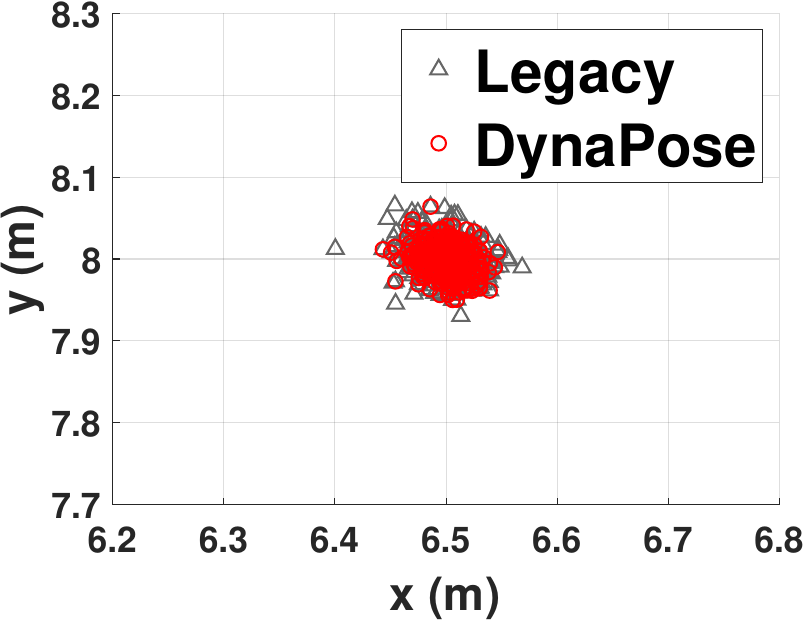}
    \hspace{-2mm}
    \includegraphics[width=0.24\textwidth]{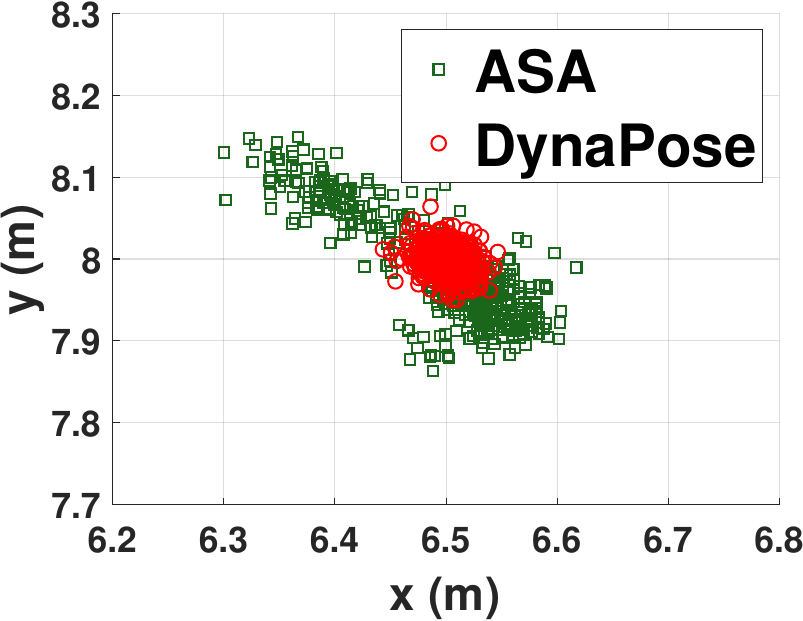}
    \label{fig:scatter_los}
    }
    \hspace{-1mm}
    \subfigure[NLOS condition]{
    \includegraphics[width=0.24\textwidth]{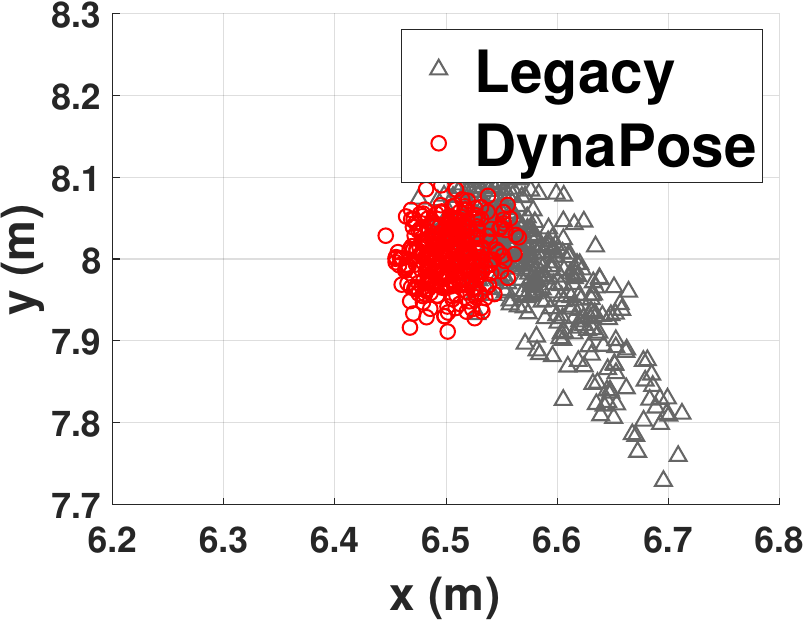}
    \hspace{-2mm}
    \includegraphics[width=0.24\textwidth]{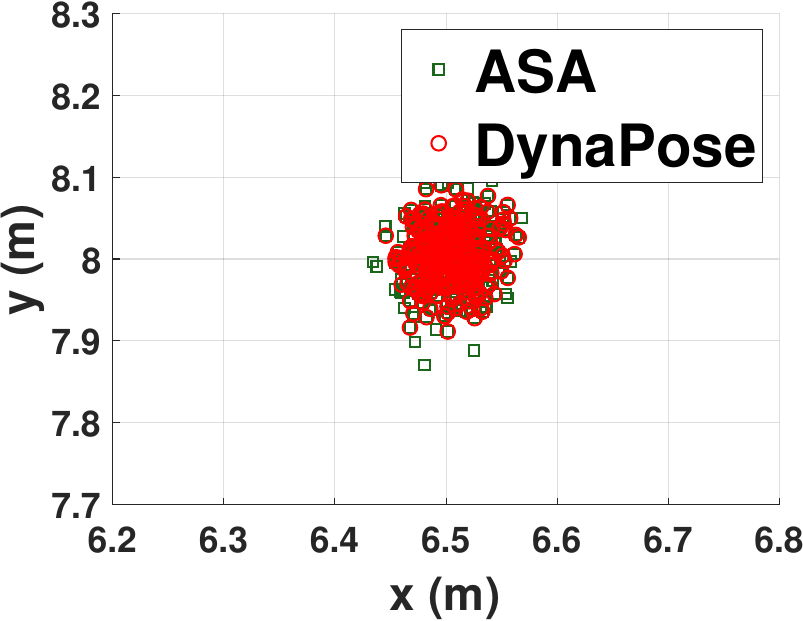}
    \label{fig:scatter_nlos}
    }
    \caption{Scatter plots of comparison schemes and {\pp} in LOS and NLOS conditions.}
    \label{fig:scatter}
\end{figure*}

\begin{figure*}[!t]
    \centering
    \subfigure[LOS condition]{
    \includegraphics[width=0.24\textwidth]{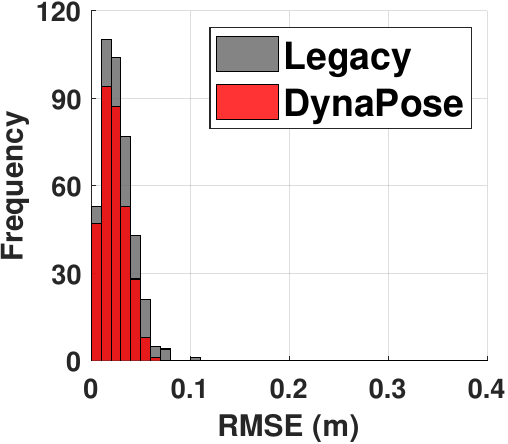}
    \hspace{-2mm}
    \includegraphics[width=0.24\textwidth]{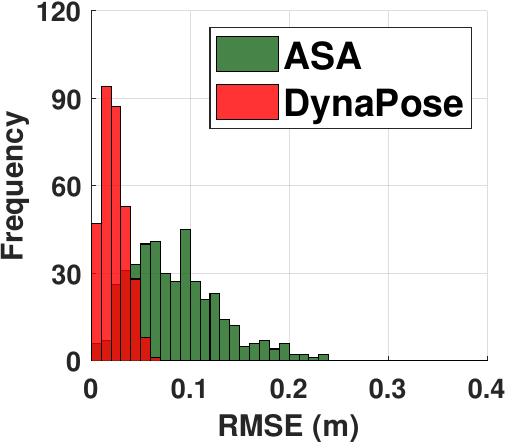}
    \label{fig:rmse_los}
    }
    \hspace{-1mm}
    \subfigure[NLOS condition]{
    \includegraphics[width=0.24\textwidth]{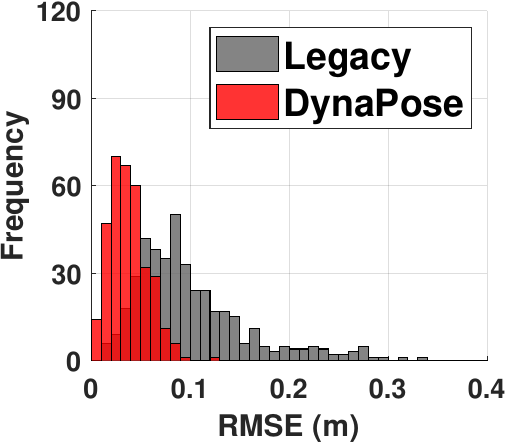}
    \hspace{-2mm}
    \includegraphics[width=0.24\textwidth]{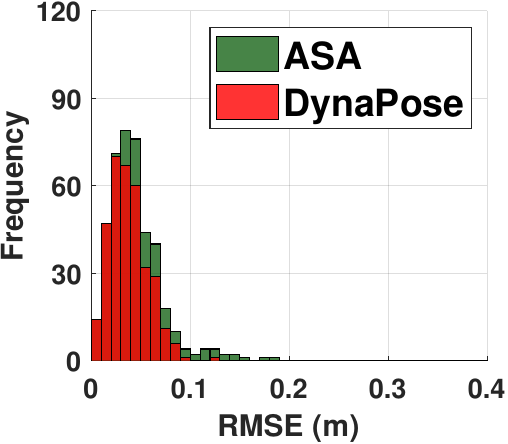}
    \label{fig:rmse_nlos}
    }
    \caption{Localization error histogram of comparison schemes and {\pp} in LOS and NLOS conditions.}
    \label{fig:rmse}
\end{figure*}

\noindent\textbf{Pose transition delay:}
{\pp} utilizes LPF to improve the accuracy of LOS/NLOS classification, and gains accuracy improvement.
However, LPF has a trade-off between accuracy and delay.
The higher the weighting factor, the longer delay.
Therefore, we evaluate the pose transition delay between LOS and NLOS condition.
To this end, we record the timing of the transition and measure the delay using application.
If {\pp} does not adopt LPF on LOS/NLOS classification, the average pose transition delay should be 200~ms, which is same as the default ranging interval of DS-TWR.
Table~\ref{tab:delay} summarizes the average transition delay and standard deviation for the possible combinations in ms.
The transition delay between HANDs takes 734.8~ms, which is the shortest among four delays, because it has no change of the position on the human body.
The remaining transitions contain pose change from/to POCKET that have more trajectory to move on the body, and consume additional 111.1~ms on average.
All pose transitions are completed within one second.

\subsection{Performance of DL-TDoA localization}
\label{subsec:performance_asa}
To measure DL-TDoA localization performance, we calculate the location of MD at (6.5, 8) from the edge of the gate access area as shown in Fig.~\ref{fig:real-world}.
Fig.~\ref{fig:scatter} and Fig.~\ref{fig:rmse} show the scatter plot of the estimated position and root mean square error~(RMSE) in the histogram, respectively.
To show the clear difference between {\pp} and comparison schemes, we illustrate each result with only one comparison scheme.
In each scenario, {\pp} is compared with a legacy and ASA using the same experiment result.
In LOS condition, both legacy and {\pp} showed robust localization results, but {\pp} has smaller standard deviation due to outlier detection as shown in Fig.~\ref{fig:scatter_los}.
On the other hand, ASA is scattered with high variance
because ASA uses only 4 anchors among 6 due to the anchor selection operation, but {\pp} leverages all TDoA values from anchors in LOS condition.
This implies that the more anchors are used, the higher localization accuracy in LOS condition due to the error compensation from multiple TDoA values.
In the NLOS condition, the scatter plot shows the degraded performance of legacy as shown in Fig.~\ref{fig:scatter_nlos}.
Since legacy utilizes all 6 anchors in computation, legacy is directly affected by NLOS anchors that contain TDoA values with the biased mean and high standard deviation. 
On the other hand, ASA and {\pp} show the robust localization performance in NLOS condition thanks to anchor selection operation eliminating poor quality of signal. 

\begin{table}[t]
\centering
    \caption{Localization performance with comparison schemes.}
    \begin{tabular}{c|c|c}
    \toprule[0.6pt]\midrule[0.3pt]
    {\textbf{Algorithm}} & \textbf{Condition} & \textbf{Mean error (cm)}  \\
    \midrule
    \multirow{2}{*}{Legacy} & LOS & 2.64 (2.10)  \\
    & NLOS & 10.01 (5.05) \\
    \midrule
    \multirow{2}{*}{ASA} & LOS & 8.45 (4.61) \\
    & NLOS & 4.43 (2.57) \\
    \midrule
    \multirow{2}{*}{{\pp}} & LOS & 2.23 (1.21) \\
    & NLOS & 3.78 (1.87) \\
    \midrule[0.3pt]\bottomrule[0.6pt]
  \end{tabular}
\label{tab:localization-error}
\end{table}

The RMSE histogram is also compared by using the experiment results.
The trends of the result are the same as the scatter plots.
In Fig.~\ref{fig:rmse_los}, {\pp} and legacy showed high localization accuracy in LOS condition, but ASA scored higher RMSE than other two algorithms due to the reduced number of anchors.
In Fig.~\ref{fig:rmse_nlos}, ASA and {\pp} showed small RMSE compared to the legacy, which demonstrates the effectiveness of anchor selection in NLOS condition. 
The overall experiment result is summarized in Table~\ref{tab:localization-error}, which shows the mean error and in the parentheses shows the standard deviation.
This implies that anchor selection is efficient in NLOS condition but not in LOS condition.
This can be also proven by comparing legacy and {\pp} that {\pp} performs anchor selection only in NLOS condition.
The result shows that {\pp} has reduced mean from 10.01~cm to 3.78~cm, and standard deviation has reduced from 5.05~cm to 1.87~cm in NLOS condition compared to legacy.
{\pp} also had improvement in LOS condition, with mean reduced from 2.64~cm to 2.23~cm and standard deviation reduced from 2.1~cm to 1.21~cm due to outlier detection.
From the experiment, we proved that {\pp} improved 62\% and 15\% DL-TDoA localization accuracy compared to legacy in NLOS and LOS conditions, respectively. 
\section{Related Work}
\label{sec:related_work}
To the best of our knowledge, this is the first work to implement real-time anchor selection and pose prediction based on UWB on commercial smartphones, including LOS/NLOS classification.

The previous work focuses on LOS/NLOS classification itself or UWB error mitigation based on the classification results, and none of the previous research has considered how to use the classification results for practical UWB application such as UTG.

\noindent\textbf{DL-TDoA localization:} We identified the LOS/NLOS condition of anchors using CIR data and performed anchor selection to discard the NLOS anchors to mitigate localization error.
Most of the papers from other researchers have a similar stream, identifying NLOS and mitigation~\cite{yang2022robust}.
Kim \textit{et al}.~\cite{kim2022uwb} 
classifies NLOS condition by feeding CIR to a pre-trained LSTM model and mitigates the distance measurement based on the output.
It once again mitigates NLOS error by extended Kalman filter~(EKF) localization.
Wen \textit{et al}.~\cite{wen2017nlos} 
analyze signal characteristics such as kurtosis and skewness and identifies not only NLOS condition but also the cause of NLOS, whether it is due to door, wall, etc, and mitigate the distance measurements.
This method reduced the ranging RMSE from 77~cm to 33~cm.
Yu \textit{et al}.~\cite{yu2018novel} 
identifies NLOS condition and mitigates through fuzzy comprehensive evaluation. 
It then performs range selection before localization, and during the process, geometric dilution of precision and channel condition is considered.
This method identified NLOS and LOS signals with 93.9\% and 92\% respectively, and localization RMSE was reduced from 106~cm to 60~cm.
Yang \textit{et al}.~\cite{yang2021novel} 
identifies NLOS condition using localized position and acceleration of IMU sensor.
It then mitigates the error by using the UWB measurements, motion trend acquired by IMU, and environmental factors about how severe the NLOS is.
It achieved 80\% localization improvement in the NLOS area.
Feng \textit{et al}.~\cite{feng2023adaptive} detects NLOS based on the support vector machine model using UWB measurements and IMU sensor.
Depending on the number of detected LOS signals, it adopts different localization algorithms.
The algorithms have similar framework, UWB/IMU sensor fusion positioning based on KF or EKF.

\noindent\textbf{Pose prediction:} We are focusing on applications such as pose prediction by combining IMU sensor data for proximity services as well as real-time LOS/NLOS classification.
Jiang~\emph{et al.}~\cite{jiang2020los} propose CNN-LSTM deep learning method for UWB LOS/NLOS classification based on CIR, which does not consider the input size of CIR samples for real-time operation on MD.
Jiang~\emph{et al.}~\cite{jiang2020denoise} focus on de-nosing the CIR data to identify NLOS signal based on CNN model, and test the effect of the de-nosing CIR method.
Wymeersch~\emph{et al.}~\cite{wymeersch2012machine} present the approach to directly mitigate ranging errors in both LOS and NLOS conditions.
Zeng~\emph{et al.}~\cite{zeng2018nlos} take consideration of the necessary size of CIRs among whole CIR samples, but does not conduct any implementation and real-world experiment.
Barral~\emph{et al.}~\cite{barral2019nlos} employ machine learning techniques to analyze several sets of real UWB measurements to identify the NLOS propagation condition, and conduct experiments with diverse machine learning algorithms.

\section{Conclusion}
\label{sec:conclusion}

In this paper, we have proposed {\pp}, a novel anchor selection pose prediction system for the proximity services based on UWB CIR and fusion of IMU sensors.
We implemented and evaluated {\pp} in a real-world environment using SM-N986B and MDEK1001.
{\pp} achieved {\cc} and {\dt} accuracy in LOS/NLOS classification and pose prediction, and improved DL-TDoA localization accuracy with 62\% in NLOS condition in real-time.


\bibliographystyle{IEEEtran}
\bibliography{sbpa2023.bib}

\end{document}